\documentclass[conference]{IEEEtran}
\IEEEoverridecommandlockouts
\usepackage{cite}
\usepackage{amsmath,amssymb,amsfonts}
\usepackage{algorithm}
\usepackage{graphicx}
\usepackage{textcomp}
\usepackage{xcolor}
\usepackage{csvsimple}
\usepackage{booktabs}
\usepackage{makecell}
\usepackage{amsmath}
\usepackage{bm}
\usepackage[noend]{algpseudocode}

\def\BibTeX{{\rm B\kern-.05em{\sc i\kern-.025em b}\kern-.08em
    T\kern-.1667em\lower.7ex\hbox{E}\kern-.125emX}}
\begin{document}

\title{EcoMLS: A Self-Adaptation Approach for Architecting Green ML-Enabled Systems
}

\author{\IEEEauthorblockN{Meghana Tedla}
\IEEEauthorblockA{\textit{Software Engineering Research Center} \\
\textit{IIIT Hyderabad, India}\\
meghana.tedla@students.iiit.ac.in}
\and
\IEEEauthorblockN{Shubham Kulkarni}
\IEEEauthorblockA{\textit{Software Engineering Research Center} \\
\textit{IIIT Hyderabad, India}\\
shubham.kulkarni@research.iiit.ac.in}
\and
\IEEEauthorblockN{Karthik Vaidhyanathan}
\IEEEauthorblockA{\textit{Software Engineering Research Center} \\
\textit{IIIT Hyderabad, India}\\
karthik.vaidhyanathan@iiit.ac.in}
}

\maketitle

\begin{abstract}
The sustainability of Machine Learning-Enabled Systems (MLS), particularly with regard to energy efficiency, is an important challenge in their development and deployment. Self-adaptation techniques, recognized for their potential in energy savings within software systems, have yet to be extensively explored in Machine Learning-Enabled Systems (MLS), where runtime uncertainties can significantly impact model performance and energy consumption. This variability, alongside the fluctuating energy demands of ML models during operation, necessitates a dynamic approach.
Addressing these challenges, we introduce EcoMLS approach, which leverages the Machine Learning Model Balancer concept to enhance the sustainability of MLS through runtime ML model switching. By adapting to monitored runtime conditions, EcoMLS optimally balances energy consumption with model confidence, demonstrating a significant advancement towards sustainable, energy-efficient machine learning solutions.
Through an object detection exemplar, we illustrate the application of EcoMLS, showcasing its ability to reduce energy consumption while maintaining high model accuracy throughout its use. This research underscores the feasibility of enhancing MLS sustainability through intelligent runtime adaptations, contributing a valuable perspective to the ongoing discourse on energy-efficient machine learning.
\end{abstract}

\section{Introduction}
The growth of Artificial Intelligence (AI) has led to significant advancements as well as environmental concerns. Machine Learning-Enabled Systems (MLS), important for developing technologies such as autonomous vehicles and smart cities, consume considerable energy and generate high carbon emissions \cite{Reuther_2019,en16155718}. Strubell et al. (2019) highlight the substantial energy required to train ML models\footnote{unless specified otherwise by the model we imply ML model in this paper}, equating the carbon footprint of training one model to that of five average cars \cite{strubell2019energy}. The drive for more accurate models increases both energy use and emissions \cite{GARCIAMARTIN201975}. Therefore, adopting sustainable AI practices is vital to balance technological advancement with environmental preservation \cite{martinez2022software,lacoste2019quantifying, parrot}.

Despite awareness of the environmental impact of training and deploying ML models, the energy efficiency of model inference—particularly in terms of its significant energy consumption and its implications for Quality of Service (QoS)—has been less addressed. Green AI initiatives have primarily focused on optimizing the training phase, with less attention given to the energy demands of inference in practical applications \cite{su13168952,greenlr,jarvenpaa2023synthesis,datacentic}. This gap highlights the need for strategies that reduce energy consumption without compromising performance and can adjust to varying operational demands. The potential of self-adaptation techniques, which balance energy efficiency with QoS, remains largely unexplored in this context \cite{saslr}. As the ICT sector's energy consumption is expected to increase, creating adaptive, energy-efficient MLS is paramount \cite{verdecchia2023systematic}. Our work seeks to bridge this gap, proposing a self-adaptive approach aiming to ensure MLS sustainability amidst environmental uncertainties.

Initial studies \cite{casimiro2021self} explored how ML systems adapt to environmental shifts.  Our further research \cite{kulkarni2023towards} introduced the Machine Learning Model Balancer, which dynamically switches between ML models to ensure optimal QoS, yet overlooked energy efficiency—a critical aspect given ML deployments' increasing energy demands and environmental impact. 
To bridge this gap, we extend the Machine Learning Model Balancer concept to include energy efficiency, aiming to create a more sustainable, adaptive MLS and introduce EcoMLS. This novel self-adaptive approach extends the Machine Learning Model Balancer concept, incorporating energy efficiency to tackle unexplored runtime uncertainties and the impact of dynamic environmental factors on model performance. EcoMLS distinguishes itself by dynamically adjusting energy consumption and model confidence through ML model switching, responding in real-time to operational conditions and request variability. Through the MAPE-K loop, it emphasizes energy efficiency alongside accuracy, involving: i) monitoring model and system energy parameters; ii) analyzing energy performance to pinpoint inefficiencies; iii) switching models based on energy-accuracy evaluations; iv) implementing strategies for sustainable MLS. EcoMLS emerges as a novel solution in self-adaptive MLS, optimizing energy consumption while aiming to maintain performance.

Our case study on object detection with EcoMLS uses the underlying ML system of our SWITCH exemplar, designed for evaluating self-adapting ML-enabled systems. This practical application confirms EcoMLS's ability to effectively manage the trade-off between energy efficiency and accuracy, showcasing its capability to be applied across a diverse set of ML models. The flexibility of EcoMLS allows it to navigate the trade-offs between maximizing confidence scores with minimal energy increases and significantly reducing energy consumption with minimal impact on model confidence. The results show that EcoMLS outperforms both naive strategies and individual models in terms of energy consumption and confidence score trade-off, demonstrating the practicality of embedding energy efficiency within self-adaptive systems. This moves us towards a future where MLS can easily meet different needs, making them sustainable for various uses.

The remainder of the paper is structured as follows: Section \ref{sec:running-example} provides a running example. Section \ref{sec:approach} introduces the EcoMLS approach. Experimentation and results from its application are in Section \ref{sec:experimentation-and-results}. Threats to validity and Related work are discussed in Section \ref{sec:threats-to-validity} and \ref{sec:related-work} respectively. Section \ref{sec:conclusion} concludes and discusses future work.

\section{Running Example}
\label{sec:running-example}
Our approach EcoMLS is exemplified through the adoption of the SWITCH~\cite{marda2024switch} exemplar's managed system and environment manager, an object detection web service. This service integrates a streamlined architecture for handling image processing requests: an {\em Image Ingestion Service} for emulating real-world asynchronous requests; an {\em Image Store} acting as a dynamic FIFO queue for incoming data; a {\em Data Preprocessor} preparing images for detection; a {\em Model Loader} for dynamic selection of machine learning models governed by underlying approach; a {\em Model Repository} housing a variety of preloaded models for quick deployment; an {\em ML Model} at the core of detection processing; a {\em Post Processor} refining detection results; and a {\em Result Storage} for temporary data retention before final transfer. This setup as shown in Figure \ref{fig:approach} mirrors services akin to Google Cloud Vision or Amazon Rekognition, underscoring the practical applicability and flexibility in managing object detection tasks with efficiency and scalability. 

In this system, we define a set of machine learning models \(M\), where each model \(m_j\) in \(M\) represents a different configuration of the YOLO algorithm \cite{redmon2016you}, varying in size and computational efficiency. These models are evaluated on two critical metrics: energy consumption, reflecting the electricity used during inference, and the confidence score, which measures the probability of accurate object identification and classification, serving as a indicator for model accuracy. Here, \(M\) includes the YOLOv5 models: YOLOv5n (nano), YOLOv5s (small), YOLOv5m (Medium), and YOLOv5l (Large), provided by Ultralytics~\cite{yolov5}, each pretrained on the COCO 2017 training dataset~\cite{coco}. The differentiation among these models primarily lies in their number of parameters, influencing both their energy consumption (\(E_j\)) and detection confidence score (\(c_j\)), where \(j\) indexes the model within \(M\).

The variants within \(M\)—ranging from YOLOv5n to YOLOv5l—exhibit a trade-off between \(c_j\) and \(E_j\). This trade-off highlights the balance required in practical applications between model accuracy, indicated by the confidence score, and operational efficiency, as represented by energy consumption. Specifically, YOLOv5n, optimized for efficiency, consumes 2 mJ of energy (\(E_1\)) with a mAP of 45.7 (\(c_1\)), while YOLOv5l, aimed at higher accuracy, consumes 16 mJ (\(E_n\)) achieving a mAP of 68.9 (\(c_n\)). This delineation showcases the balance between computational demand and detection accuracy.

In operational terms, the system processes a stream of image requests \(i\), selecting a model \(m_j\) for each \(i\), based on optimizing the balance between energy efficiency (\(E_i\)) and detection confidence (\(c_i\)). Our approach, EcoMLS aims to dynamically switch between the models to maintain an effective balance between \(E_i\) and \(c_i\), enhancing both the sustainability and the performance of the object detection service.

\section{Approach}
\label{sec:approach}
\begin{figure}[!t]
    \centering
    \includegraphics[width=\columnwidth]{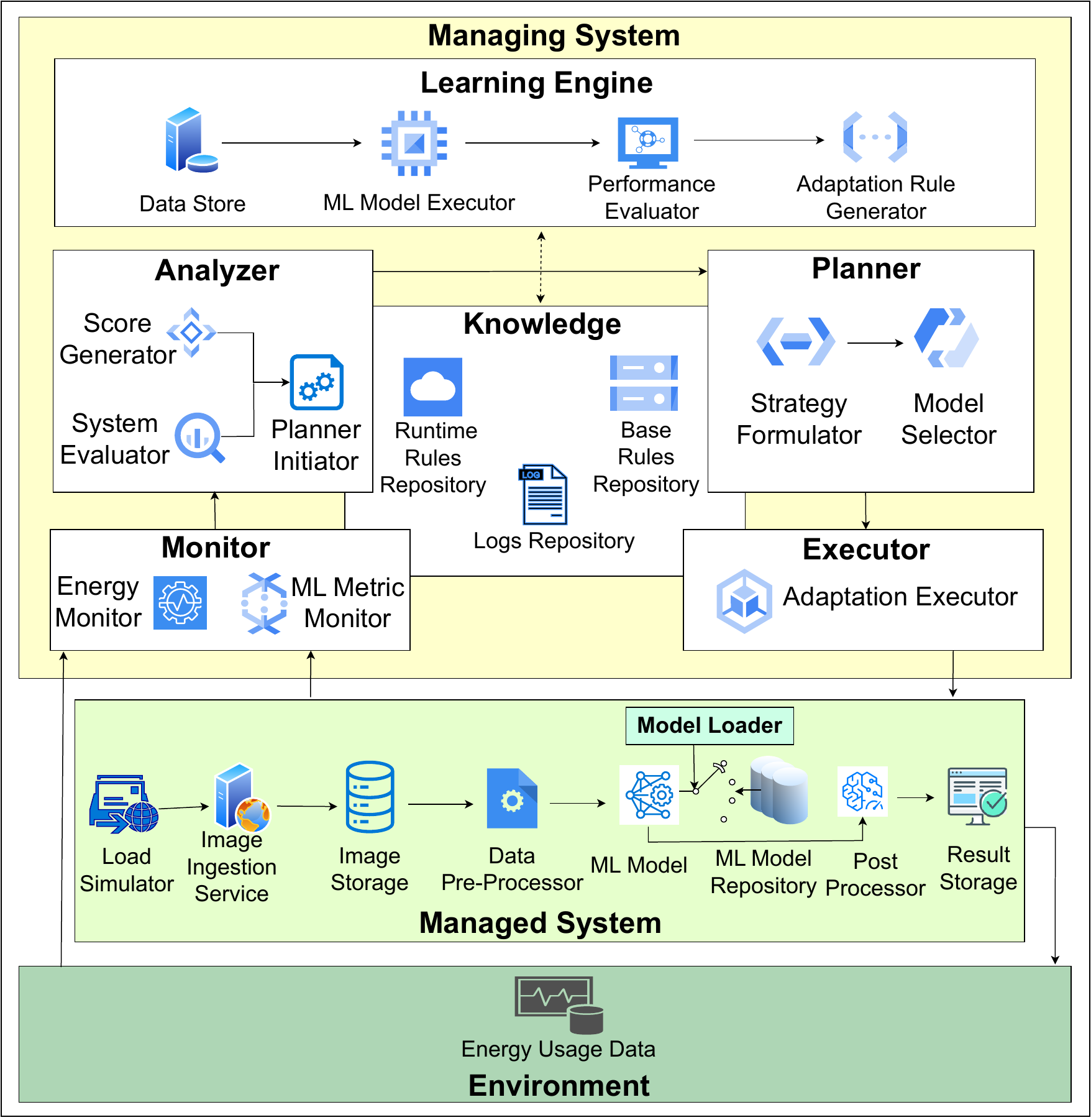}
    \caption{Architecture of EcoMLS}
    \label{fig:approach}
    \vspace*{-0.5cm}
\end{figure}

Our EcoMLS approach focuses on improving machine learning-enabled systems with an eye on sustainability and performance. The first step is to explore the Learning Engine, as shown in Figure \ref{fig:approach}.

\subsection{Learning Engine}
The Learning Engine (LE) within EcoMLS as shown in Figure \ref{fig:approach} is structured to evaluate and adapt a diverse set of machine learning models \(M\), to enhance both sustainability and performance within the MLS framework. This engine is built to work with a variety of models and datasets, offering a robust foundation for dynamic adaptation across different scenarios. Within the LE, the {\em Data Store (DS)} serves as the repository for both the evaluation dataset and the models \(M = \{m_j\}\). These models are assessed using the evaluation dataset to determine their performance. 
The \emph{ML Model Executor (MLE)} deploys the models against the evaluation dataset. For each model \(m_j \in M\), it records key performance metrics across all evaluated requests in the dataset. These metrics include energy consumption (\(E_j\)), confidence score (\(c_j\)), and processing time (\(t_{mp,j}\)). Logging these metrics is essential for evaluating each model's operational efficiency and prediction/ detection accuracy.

For performance analysis, the \emph{Performance Evaluator (PE)} constructs a performance matrix \(\mathbf{P}_j\) for each model \(m_j\). Suppose the evaluation dataset contains \(r\) requests. In that case, \(\mathbf{P}_j\) will have \(r\) rows, with each row corresponding to a request in the dataset and columns representing the key metrics: energy consumption (\(E_j\)), confidence score (\(c_j\)), and processing time (\(t_{mp,j}\)). This component enables a thorough comparison of model performance.

Building on the insights gained from each \(\mathbf{P}_j\), the \emph{Adaptation Rule Generator} synthesizes a set of base adaptation rules. It does so by averaging the metrics in \(\mathbf{P}_j\) for each model \(m_j\), resulting in aggregated values of minimum energy (\(E_{min,j}\)), maximum energy (\(E_{max,j}\)), and average confidence score (\(C_{avg,j}\)) across all requests. These aggregated values are then represented in a matrix \(\mathbf{A}\):
\[
\mathbf{A} = \begin{bmatrix}
E_{min,1} & E_{max,1} & C_{avg,1} \\
\vdots & \vdots & \vdots \\
E_{min,n} & E_{max,n} & C_{avg,n}
\end{bmatrix}
\]
In the matrix \(\mathbf{A}\), each row corresponds to a distinct model \(m_j\) within the set \(M\), where \(j\) ranges from 1 to \(n\), indicating the sequence of models considered in the evaluation. The columns \(E_{min,j}\), \(E_{max,j}\), and \(C_{avg,j}\) represent, respectively, the minimum energy consumption, maximum energy consumption, and average confidence score calculated for each model \(m_j\) across all requests in the evaluation dataset. The matrix \(\mathbf{A}\) gives a clear performance comparison of each model, guiding the selection based on efficiency and accuracy. In the context of our running example \ref{sec:running-example}, \(m_j\) refers to YOLOv5 variants~\cite{yolov5} -YOLOv5n, YOLOv5s, YOLOv5m, and YOLOv5l, evaluated against the COCO Test 2017~\cite{coco} to benchmark performance. The requests here are images, illustrating how the LE applies to the object detection use cases.

\subsection{MAPE-K Loop}
\subsubsection{Knowledge}
In the EcoMLS's MAPE-K (Monitor, Analyze, Plan, Execute - Knowledge) framework as depicted in Figure \ref{fig:approach}, the Knowledge component (\(K\)) integrates three key repositories essential for the system's adaptive behavior: the \emph{Log Repository}, the \emph{Base Rules Repository}, and the \emph{Runtime Rules Repository}. These repositories collectively maintain the data necessary for informed decision-making and dynamic adaptation of the system.
The \emph{Log Repository} stores real-time logs of each processed request, including energy consumption (\(E_j\)), confidence score (\(c_j\)), request processing time (\(t_{sys}\)), and detection outcomes such as the number of detected bounding boxes (\(b\)) for running example explained in Section \ref{sec:running-example}. This repository is continuously updated, ensuring a comprehensive record of system performance metrics.
The \emph{Base Rules Repository} is directly derived from the Learning Engine's output, specifically referencing the matrix \(\mathbf{A}\) previously discussed. This matrix, generated for the initial evaluation of models against the evaluation dataset, establishes a baseline for energy efficiency (\(E_{min,j}\) and \(E_{max,j}\)) and confidence score (\(C_{avg,j}\)) for each model \(m_j\). It serves as the foundation for model selection and system adaptation during the initial operational phase.
Post the initial model evaluations, the \emph{Runtime Rules Repository} (\( \mathbf{B} \)) updates real-time performance metrics for each model \( m_j \) as new requests are processed. The structure of this repository is formalized as a matrix that dynamically reflects changes in model performance and energy usage:
\[
\mathbf{B} = \begin{bmatrix}
E_{min,1} & E_{max,1} & E_{latest,1} & C_{avg,1} \\
E_{min,2} & E_{max,2} & E_{latest,2} & C_{avg,2} \\
\vdots & \vdots & \vdots & \vdots \\
E_{min,n} & E_{max,n} & E_{latest,n} & C_{avg,n}  \\
\end{bmatrix}
\]
In the matrix \( \mathbf{B} \), each row corresponds to a model \( m_j \) in the set \( M \), with columns for minimum and maximum energy consumption (\( E_{min,j} \) and \( E_{max,j} \)), observed from initial evaluations. \( E_{latest,j} \) represents the most recent energy consumption for model \( m_j \). \( C_{avg,j} \) is the running average of confidence score for model \( m_j \), calculated over the latest \( k \) processed requests. This repository's dynamic nature allows for continuous refinement of adaptation rules based on the most current data, facilitating an adaptive and responsive system capable of optimizing model selection in real time for enhanced performance and energy efficiency. It allows the Knowledge component to provide real-time insights into model efficacy, facilitating adaptive system responses to changing operational conditions.

\subsubsection{Monitor}
The Monitor in the MAPE-K framework is divided into two specialized components: the \emph{Energy Monitor} and the \emph{ML Metric Monitor}, each focusing on distinct aspects of system performance monitoring.
The \emph{Energy Monitor} specifically tracks the energy consumption metrics (\(E_{m}\)) of the system, including the average energy consumption for the last \(k\) processed requests (\(\bar{E}_{k}\)), providing a near-real-time insight into the system's power efficiency.
Also, the \emph{ML Metric Monitor} focuses on the detection performance metrics of the currently active model (\(m_{current}\)), including the average confidence score (\(\bar{C}_{k}\)) across the last \(k\) images processed by the model, where \(k\) represents a variable number of recent requests considered for generating a moving average. This component not only monitors confidence score but also logs additional detection outcomes like the count of bounding boxes into the \emph{Log Repository} in Knowledge, facilitating a thorough performance review. Together, these components compile a continuous log of energy and ML metrics. The averaged data from the latest \(k\) observations is continuously sent to the Analyzer for further examination and to assess the need for adaptation. This steady stream of up-to-date information is essential for the adaptive functions of the EcoMLS approach.
\subsubsection{Analyzer}
The Analyzer component evaluates the real-time data from the Monitor and decides system adaptations. Its functionality is segmented into three components: \emph{System Evaluator}, \emph{Score Generator}, and \emph{Planner Initiator}, each contributing to the adaptive decision-making process. \emph{System Evaluator} analyses the performance metrics provided by the Monitor, specifically focusing on the averaged energy consumption (\(\bar{E}_k\)) and confidence score (\(\bar{C}_k\)) for the last \(k\) processed requests. It ensures the \emph{Runtime Rule Repository} (\(\mathbf{B}\)) is up-to-date with the latest performance data before any adaptation decisions are made.
Following the evaluation, the \emph{Score Generator} calculates a performance score (\(\text{Score}_{m}\)) for the currently active model (\(m_{current}\)), using the formula \(E_j \times (1 - c_j)\). The aim is to minimize this score, where a lower score indicates a more efficient balance between energy consumption and confidence score. Before any updates are made to the \emph{Runtime Rules Repository} (\(\mathbf{B}\)), the Analyzer provides the Planner with the current state of \(\mathbf{B}\), allowing it to devise a strategic adaptation plan based on the most recent data. This step ensures that the Planner's decisions are informed by the latest model performance insights. The \emph{Planner Initiator} then triggers the Planner to evaluate whether a model switch or any other adaptation action is necessary. If the performance score (\( \text{Score}_{m} \)) for the currently active model (\( m_{current} \)) exceeds the threshold defined by its corresponding metrics in the \emph{Runtime Rule Repository} (\( \mathbf{B} \)), using \( E_{avg} \), where \(E_{avg}\) is the average of \(E_{min}\) and \(E_{max}\), and \( C_{avg} \) from \( \mathbf{B} \), it indicates inefficiency or ineffectiveness of the current model in use. This discrepancy triggers the system to initiate an adaptation, leveraging the latest data from \( \mathbf{B} \) to optimize the system's performance.
This approach ensures the Analyzer updates the Knowledge with the latest metrics and works with the Planner for prompt adaptations. This collaboration enables dynamic, informed decision-making, improving the EcoMLS system's adaptability and optimization.

\subsubsection{Planner}
The Planner component strategizes adaptations based on insights from the Analyzer, specifically addressing energy efficiency and confidence score through model selection under two scenarios as explained in Algorithm \ref{algo:Planner}. The algorithm navigates decision-making under two primary scenarios—high energy and low confidence—leveraging a unified score metric for model selection (lines 11-16). The score (\textit{score}\_{m}[i]), calculated for each model, combines energy efficiency and confidence inversely (line 10), guiding the selection process. In high energy scenarios (\(E_{m_{current}} > E_{avg,{m_{current}}}\)), models less energy-intensive than the average are considered, selecting the one with the lowest score to ensure energy efficiency (lines 11-13). Conversely, for low confidence (\(C_{m_{current}} < C_{avg,{m_{current}}}\)), models that promise confidence improvement without excessive energy use are preferred, again choosing the lowest score to balance confidence and efficiency (lines 14-16). The algorithm initiates with an $\epsilon$-greedy approach (line 6), randomly selecting an action if below a certain threshold $\epsilon$, otherwise it evaluates each model's score based on their energy and confidence metrics (lines 7-16). The executor performs model switching action (line 17), aiming to optimize the system for either energy efficiency or confidence enhancement as per the current operational requirement.

\begin{algorithm}
\caption{Planner: Algorithm for Model Selection}\label{algo:Planner}
    \begin{algorithmic}[1]
        \Procedure{Formulator}
        {$m_{current},E_{m_{current}}, C_{m_{current}}$}
        \Comment {Input: Current model \(m_{current}\), its energy consumption \(E_{m_{current}}\), and confidence score \(C_{m_{current}}\)}
        \State $A \gets \{NA : -1, m_1:1, m_2:2, m_3:3, m_4:4\}$
        \State \textit{$\text{p} \gets \text{random(0, 1)}$}
        \If{\( p < \epsilon \)}\\
            \Comment{Perform exploration}
            \State $action \gets randint(1, 4)$ 
        \Else \\
            \Comment{Perform exploitation based on the runtime context}
            \For{$i$ in $1$ to $4$}
                \State $\textit{score}_{m}[i] = \min(E_{\textit{avg},{m_i}}, E_{\textit{latest},{m_i}})\times (1-C_{\textit{avg},{m_i}})$ 
            \EndFor 
            \If{\textit{$\text{$E_{m_{current}}$} > \text{$E_{avg,{m_{current}}}$}$}} \\
               \Comment{Aim for energy efficiency improvement}
                \State $\textit{model} = \textit{$argmin_{i \in [1, 4], \text{s.t., } E_{\textit{avg},{m_i}} < \text{$E_{m}$}}$} \textit{ score}_{m}[i]$
            \Else \\
                \Comment {Aim to improve confidence}
                \State $\textit{model} = \textit{$argmin_{i \in [1, 4], \text{s.t., } C_{\textit{avg},{m_i}} > \text{$C_{m}$}}$} \textit{ score}_{m}[i]$
            \EndIf
        \EndIf
        \State \textit{$\text{action} \gets \text{A[model]}$}      
         \State \textbf{return }  $action$
        \EndProcedure
    \end{algorithmic}
\end{algorithm}

\subsubsection{Executor}
The Executor is responsible for actualizing the adaptation strategy decided by the Planner.
Upon receiving a directive for a model switch from the Planner, the \emph{Adaptation Executor} activates the selected model (\(m_{best}\)). In the absence of such an instruction, it continues with the currently active model (\(m_{current}\)).
This adaptability is crucial for sustaining the system's efficiency and accuracy in real-time environments.
In this discussion, 'requests' and 'metrics' are generalized, but for our running example as explained in section \ref{sec:running-example}, they refer to images and detection boxes in object detection. This illustrates the approach's versatility, applicable to a wide range of ML tasks, with further details and results explored in the following section.

\section{Experimentation and Results}
\label{sec:experimentation-and-results}
\begin{figure*}[t]
    \centering
    \centerline{\includegraphics[width=\textwidth, height=4.5in]{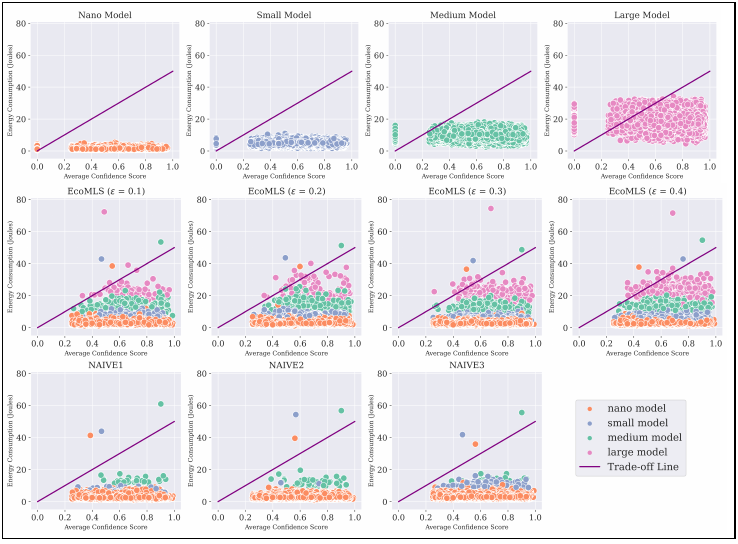}}
    \caption{Trade-off between energy consumption and the average confidence score of individual models (first row), EcoMLS with varying $\epsilon$ (second row), and naive baselines (third row).}
    \label{fig:EvC}
    \vspace*{-0.5cm}
\end{figure*}

The objective of our evaluation is to assess the effectiveness and efficiency of the approach by answering:

\noindent \textbf{RQ1.} How does the EcoMLS perform in comparison to other naive baselines and non-adaptive solutions in MLS?

\noindent \textbf{RQ2.} What is the effectiveness of EcoMLS in balancing objectives trade-offs within ML-Enabled Systems (MLS)?

\noindent \textbf{RQ3.} How effective is EcoMLS in managing energy and time efficiency in MLS environments, including the specific energy and time impacts of the EcoMLS adaptation process?

In the remainder of this section, we first describe our experimental setup used for the evaluation of the approach, followed by a discussion of the evaluation questions informed by our results.

% Switching
\begin{figure*}[ht]
    \centering
    \centerline{\includegraphics[width=\textwidth, height=1.7in]{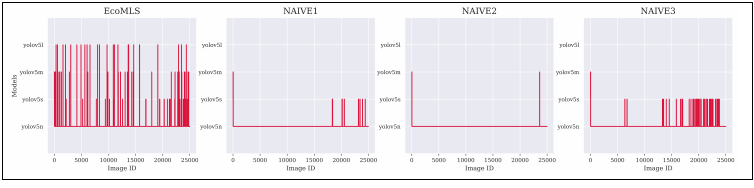}}
    \caption{Model switching: Naive baselines Vs. EcoMLS}
    \label{fig:switching}
    \vspace*{-0.5cm}
\end{figure*}

\subsection{Experimental Setup}
To evaluate the EcoMLS approach, our experimental setup is adopted from the SWITCH~\cite{marda2024switch} exemplar. It employs an object detection system, as detailed in Section \ref{sec:running-example}, utilizing YOLOv5 models and FastAPI simulating requests using the FIFA98 World Cup log dataset, processing 25,000 image requests. For evaluation, as detailed in our approach, the COCO 2017 test dataset is utilized. For the testing phase, which includes performance results, the COCO 2017 unlabelled dataset is employed. The YOLOv5 models (YOLOv5n, YOLOv5s, YOLOv5m, YOLOv5l) used are pre-trained by Ultralytics on the COCO 2017 training dataset. The experiments were conducted on a system equipped with an Intel Core i7-11370H processor, NVIDIA GeForce RTX 3050 Ti 4GB Graphics, 16GB DDR4 3200MHz SDRAM, and developed using Python 3.11. To measure energy consumption, we utilized pyRAPL\footnote{https://pypi.org/project/pyRAPL/. Latest version released on Dec 19, 2019}, a Python package specifically designed for assessing the energy consumption and power usage of software applications running on Intel processors. The EcoMLS framework's evaluation included varying the $\epsilon$ value to analyze its impact on balancing exploration and exploitation. We compared EcoMLS's adaptive model selection with individual YOLOv5 model performances and three naive strategies: (1) using fixed knowledge for model switching (naive 1), (2) updating knowledge based on average confidence (naive 2), and (3) incorporating dynamic updates of both confidence and energy metrics in knowledge (naive 3). Our approach builds on naive 3 by adding an $\epsilon$-greedy mechanism, enhancing exploration through probability. The complete specifics of our implementation, parameters and results can be found here\footnote{https://github.com/sa4s-serc/EcoMLS}.

\begin{table*}[!t]
    \centering
    \begin{tabular}{|l|c|c|c|c|c|c|c|c|c|}
        \hline
        \textbf{Approach name} & \textbf{$\mathbf{C_{avg}}$} & $\mathbf{E}_{\textbf{avg}}$ & $\mathbf{E}_{\textbf{monitor}}$ & $\mathbf{E}_{\textbf{analyzer}}$ & $\mathbf{E}_{\textbf{planner}}$ & $\mathbf{E}_{\textbf{executor}}$ & $\mathbf{E}_{\textbf{mape-k}}$ & $\mathbf{E}_{\textbf{avg}}+\mathbf{E}_{\textbf{mape-k}}$ & \textbf{No. of Switches}\\
        \hline
        \textbf{nano} & 0.536 & 1.61 & - & - & - & - & - & 1.61 & 0\\
        \hline
        \textbf{small} & 0.611 & 4.327 & - & - & - & - & - & 4.327 & 0\\
        \hline
        \textbf{medium} & 0.652 & 8.918 & - & - & - & - & - & 8.918 & 0\\
        \hline
        \textbf{large} & 0.675 & 17.705 & - & - & - & - & - & 17.705 & 0\\
        \hline
        \textbf{EcoMLS ($\epsilon = 0.1$)} & 0.61 & 2.762 & 1.284 & 0.001 & 0.001 & 0.0 & 1.285 & 4.047 & 160\\
        \hline
        \textbf{EcoMLS ($\epsilon = 0.2$)} & 0.612 & 3.044 & 1.166 & 0.001 & 0.001 & 0.0 & 1.168 & 4.212 & 324\\
        \hline
        \textbf{EcoMLS ($\epsilon = 0.3$)} & 0.613 & 2.912 & 1.035 & 0.001 & 0.001 & 0.0 & 1.037 & 3.949 & 313\\
        \hline
        \textbf{EcoMLS ($\epsilon = 0.4$)} & 0.616 & 3.564 & 0.959 & 0.001 & 0.001 & 0.001 & 0.961 & 4.525 & 676\\
        \hline
        \textbf{NAIVE1} & 0.609 & 2.47 & 1.226 & 0.001 & 0.001 & 0.0 & 1.228 & 3.697 & 20\\
        \hline
        \textbf{NAIVE2} & 0.609 & 2.399 & 1.21 & 0.001 & 0.001 & 0.0 & 1.213 & 3.612 & 5\\
        \hline
        \textbf{NAIVE3} & 0.609 & 3.319 & 1.658 & 0.001 & 0.002 & 0.0 & 1.661 & 4.98 & 106\\
        \hline
    \end{tabular}
    \vspace{3pt}
    \caption{Comparison of energy metrics and confidence scores across different approaches}
    \label{tab:Greens table}
    \vspace*{-0.5cm}
\end{table*}

\subsection{Results}
\textit{RQ1: How does the EcoMLS perform in comparison to other naive baselines and non-adaptive solutions in MLS?}

\hfill \break
We examine EcoMLS's effectiveness by comparing its performance to state-of-the-art baselines and non-adaptive models, focusing on energy consumption and confidence scores balance. Table \ref{tab:Approach Scores} showcases that within the score range of 0-1, EcoMLS ($\epsilon = 0.1$) processes 13,040 images, showcasing its capability to frequently achieve lower scores, indicative of a balanced trade-off between energy and accuracy. In contrast, the nano model processes 20,265 images within the same score range, suggesting a tendency towards lower energy consumption but at a compromise of average confidence ($C_{avg} = 0.536$), as highlighted in Table \ref{tab:Greens table}.

\begin{figure}[!t]
    \centering
    \centerline{\includegraphics[width=3in, height=2in]{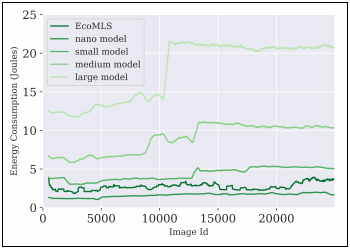}}
    \caption{Trend of energy consumption with processed image requests}
    \label{fig:EvImage}
    \vspace*{-0.5cm}
\end{figure}

The large model, which is the most accurate but also the most energy-consuming, achieves an average confidence ($C_{avg}$) = 0.675 with a high energy consumption of 17.705. EcoMLS, with $\epsilon = 0.1$, strikes a balance with average confidence ($C_{avg}$) = 0.61 and significantly lower average energy consumption of 2.762, showcasing an efficient trade-off with a 14\% improvement in confidence over the nano model and an 84\% reduction in energy consumption compared to the large model as observed from Figure \ref{fig:EvImage}. EcoMLS's adaptive strategy aims to select the most energy-efficient models without compromising confidence. Unlike no switching, EcoMLS consistently ensures detections fall below a reference trade-off line, avoiding low-confidence results and effectively using the $\epsilon$-greedy mechanism for adaptation. EcoMLS's adaptability is further highlighted by its frequency of model switching, contrasting with the minimal switching in baseline models due to their static nature. EcoMLS engages in more frequent model changes as shown in Figure \ref{fig:switching}, evidenced by 160 switches with $\epsilon = 0.1$, indicating an understanding of runtime conditions. EcoMLS outperforms traditional models by smartly balancing energy use and accuracy, using adaptive strategies like model switching and $\epsilon$-greedy exploration. It slightly exceeds the nano model in energy consumption, depicted in Figure \ref{fig:EvImage}, but compensates with higher confidence levels. Compared to all other models and approaches, EcoMLS achieves better performance scores, effectively managing the trade-offs between energy and accuracy. This underscores EcoMLS's superior efficiency and its ability to deliver high-quality outcomes while using resources optimally.
\\

\textit{RQ2. What is the effectiveness of EcoMLS in balancing objective trade-offs within ML-Enabled Systems (MLS)?}

\hfill \break
We assess the balance EcoMLS achieves between energy consumption and model confidence in MLS, using the performance score $Score_m = E_j \times (1 - C_j)$, with a lower score indicating a more favorable balance. According to Table \ref{tab:Greens table}, EcoMLS\footnote{unless specified otherwise by EcoMLS we imply EcoMLS with $\epsilon = 0.1$}, particularly with $\epsilon = 0.1$, optimally balances energy and confidence, exhibiting an average energy consumption ($E_{avg}$) of 2.762 and an average confidence score ($C_{avg}$) of 0.61 (Table \ref{tab:Greens table}). Incremental analysis from $\epsilon = 0.1$ to $0.4$ reveals a 29\% rise in $E_{avg}$ for a mere 1\% improvement in $C_{avg}$, justifying $\epsilon = 0.1$ as the optimal setting due to the diminishing returns of higher exploration.

The effectiveness of EcoMLS in managing trade-offs is further evidenced by plotting energy versus confidence scores as shown in Figure \ref{fig:EvC}, with a reference line connecting the lowest and highest energy and confidence values across all models. EcoMLS's results predominantly fall below this line, especially at $\epsilon = 0.1$, affirming its efficiency in maintaining a balance between energy use and accuracy, despite the tendency for more detections to cross the line as $\epsilon$ increases, indicating more exploratory actions. EcoMLS demonstrates effective trade-off management between energy efficiency and model confidence, with $\epsilon = 0.1$ proving to be the most effective. This conclusion is underpinned by detailed quantitative analysis, illustrating EcoMLS's capability to optimize the exploration-exploitation trade-off while ensuring sustainable MLS operation.
\\

\begin{table}[!b]
    \centering
    \begin{tabular}{|l|c|c|c|c|c|c|c|c|c|c|}
    \hline
    \textbf{Name} & \multicolumn{7}{c|}{\textbf{score}}\\
    \hline
    \textbf{} & \textbf{0-1} & \textbf{1-2} & \textbf{2-3} & \textbf{3-4} & \textbf{4-5} & \textbf{5-6} & \textbf{6-7} \\
    \hline
    \textbf{nano} & 20265 & 4480 & 237 & 16 & 1 & 0 & 0\\
    \hline
    \textbf{small} & 5301 & 11870 & 6139 & 1258 & 290 & 102 & 31\\
    \hline
    \textbf{medium} & 2394 & 4447 & 6018 & 5319 & 3670 & 1870 & 706\\
    \hline
    \textbf{large} & 645 & 2398 & 2304 & 2691 & 3204 & 3148 & 2693\\
    \hline
    \textbf{EcoMLS} & 13040 & 10740 & 841 & 180 & 61 & 37 & 31\\
    \hline
    \textbf{NAIVE1} & 14301 & 9996 & 562 & 96 & 27 & 9 & 4\\
    \hline
    \textbf{NAIVE2} & 14860 & 9560 & 491 & 62 & 12 & 4 & 3\\
    \hline
    \textbf{NAIVE3} & 9188 & 12408 & 2698 & 492 & 156 & 43 & 9\\
    \hline
    \end{tabular}
    \vspace{3pt}
    \caption{Model score frequency table}
    \label{tab:Approach Scores}
    \vspace*{-0.5cm}
\end{table}

% E v ImageID

\textit{RQ3: How effective is EcoMLS in managing energy and time efficiency in MLS environments, including the specific energy and time impacts of the EcoMLS adaptation process?}

\hfill \break
In RQ3, we evaluate the energy efficiency of our adaptive approach, focusing on the MAPE-K technique. The analysis shows that the MAPE-K loop's additional energy consumption is minimal compared to traditional models. Specifically, integrating the MAPE-K loop, EcoMLS consumes 77.14\% less energy than the 'large' model, 54.62\% less than 'medium', and 6.46\% less than 'small' as shown in Table \ref{tab:Greens table}. The MAPE-K loop itself only adds a negligible time overhead of 0.007 seconds for each adaptation. EcoMLS continuously monitors system performance, with the analyzer evaluating the need for adaptation every second. This process triggers the planner and executor as needed, ensuring optimal system operation with minimal delay. For instance, with $\epsilon = 0.1$, EcoMLS's energy consumption for model operations is 2.762 Joules, with the MAPE-K loop contributing an additional 1.285 Joules, for a total of 4.047 Joules. This efficiency persists across various $\epsilon$ settings, demonstrating that while the adaptive mechanism introduces a slight energy cost for adaptations, it significantly lowers overall energy usage compared to static models, without compromising performance. With 160 switches at $\epsilon = 0.1$, EcoMLS showcases effective adaptability and operational efficiency, adeptly balancing energy use and accuracy. These values, measured in Joules by PyRAPL, indicate the precise energy impact of the MAPE-K loop, continuous monitoring and adaptation process. This evaluation shows EcoMLS's capability to maintain high performance with energy efficiency.

\section{Threats to Validity}
\label{sec:threats-to-validity}
Threats to {\em external validity} concerns the focus on a single type of task and a select group of machine learning models, and by limiting our examination to the inference phase. To address the first challenge, we chose a range of YOLOv5 models, varying in complexity (from YOLOv5n to YOLOv5l), and used diverse datasets, including COCO 2017. This strategy aimed to cover different visual data types and model sizes. However, our decision to focus only on the inference phase, without considering the full lifecycle of machine learning models like training and tuning, was intentional. This choice was made to study energy consumption during inference specifically, acknowledging its narrow scope in reflecting the entire machine learning process.

A threat to the {\em internal validity} could be the impact of varying hardware conditions like temperature changes on the results of the evaluations. To tackle this, we implemented a 24-hour sleep period before each test to stabilize hardware conditions and performed a warm-up run to maintain consistency throughout our experiments. The threats {\em construct validity} could be constituted by the accuracy of our energy consumption measurements. To mitigate this, we utilized the pyRAPL library in Python (pyRAPL makes use of the RAPL library provided by Intel) which is a well-known library for measuring energy consumption. Concerning {\em conclusion validity}, the main threat is the potential low statistical power of our tests, which we addressed by conducting multiple experiments across different settings and conditions. Additionally, we took precautions to minimize the impact of extraneous variables, such as background tasks, on our energy consumption measurements by ensuring a clean experimental environment.

\section{Related Work}
\label{sec:related-work}

The field of Green AI has attracted substantial research interest, focusing on energy efficiency in Machine Learning (ML) systems. Verdecchia et al. \cite{verdecchia2023systematic} reviewed 98 studies, highlighting a dominant emphasis on energy efficiency mechanisms. Despite this, the practical application of these findings, particularly in making ML-enabled systems sustainable, has been limited. Studies like those by Jarvenpaa et al. \cite{jarvenpaa2023synthesis} propose tactics for sustainability, mainly at the design stage, covering data-centric methods \cite{verdecchia2022data} and optimization of algorithms and models. However, the aspect of runtime energy efficiency has been relatively underexplored. The concept of self-adaptation in software, originating from IBM's autonomic computing \cite{wong2022self}, has evolved to include Machine Learning Systems (MLS). Research in this area \cite{casimiro2021self, muccini2021software, moghaddam2018self} categorizes self-adaptation into software design approaches (SDA) and system engineering approaches (SEA), focusing on design-time solutions. Recent studies \cite{gerostathopoulos2022expressing, kulkarni2023towards} explore enhancing adaptability at runtime, including unsupervised learning and model switching. Yet, these approaches often limit systems to pre-set configurations, as noted by Tundo et al. \cite{tundo2023energy}.

Our approach diverges by offering a dynamic, self-adaptive framework that addresses runtime uncertainties in ML-enabled systems. It employs a learning algorithm to monitor and adapt based on historical data, allowing for real-time model switching. This adaptive planning considers incoming request contexts and model performance indicators, optimizing for accuracy, service quality and minimal energy consumption. Our work fills the gap in runtime energy efficiency and contributes to the development of self-adaptive, sustainable ML-enabled systems.

\section{Conclusions and Future Work}
\label{sec:conclusion}
This paper introduces EcoMLS, an innovative strategy designed to improve the sustainability of Machine Learning-Enabled Systems (MLS) by employing dynamic runtime model switching to enhance energy efficiency while ensuring model confidence is not compromised. Our evaluations reveal that EcoMLS significantly boosts energy efficiency in MLS by adapting model selection in response to real-time conditions, thereby maintaining high model accuracy and substantially lowering energy consumption. Through the EcoMLS approach, which utilizes the Machine Learning Model Balancer concept within a MAPE-K loop, we address environmental uncertainties linked to MLS deployment, focusing on reducing energy consumption in runtime operations. Currently demonstrated within the computer vision domain through object detection, future works will extend EcoMLS's applicability to Natural Language Processing (NLP), autonomous systems, and beyond, exploring its potential across a broader spectrum of ML applications.
Looking ahead, we intend to investigate EcoMLS's applicability in sustainable computing, particularly its integration with edge computing and lightweight AI models. This exploration will encompass the application of software engineering methodologies attuned to the sustainability dimensions of software systems, covering economic, social, environmental, and technical aspects. Our objective is to evolve EcoMLS into a tool that facilitates sustainability-aware decision-making, enabling architects, developers and businesses to create greener, more sustainable ML-Enabled systems. In summary, EcoMLS marks a advancement towards achieving energy-efficient and sustainable ML-Enabled systems. By effectively balancing energy consumption with model confidence, EcoMLS seeks to set a new standard towards greener AI. Future efforts will focus on extending its impact, with EcoMLS continuing to lead innovation in an environmentally conscious way.

\bibliographystyle{IEEEtran}

\bibliography{bib}

\end{document}